\documentstyle[12pt]{article}

\newcommand{\nc}{\newcommand}
\nc{\beq}{\begin{equation}}
\nc{\eeq}{\end{equation}}
\nc{\beqa}{\begin{eqnarray}}
\nc{\eeqa}{\end{eqnarray}}

\input epsf
\newwrite\ffile\global\newcount\figno \global\figno=1

\def\writedef#1{}
\def\figin{\epsfcheck\figin}\def\figins{\epsfcheck\figins}
\def\epsfcheck{\ifx\epsfbox\UnDeFiNeD
\message{(NO epsf.tex, FIGURES WILL BE IGNORED)}
\gdef\figin##1{\vskip2in}\gdef\figins##1{\hskip.5in}% blank space instead
\else\message{(FIGURES WILL BE INCLUDED)}%
\gdef\figin##1{##1}\gdef\figins##1{##1}\fi}
\def\figinsert{}
\def\ifig#1#2#3{\xdef#1{fig.~\the\figno}
\writedef{#1\leftbracket fig.\noexpand~\the\figno}%
\figinsert\figin{\centerline{#3}}\medskip\centerline{\vbox{\baselineskip12pt
\advance\hsize by -1truein\center\footnotesize{  Fig.~\the\figno.} #2}}
\bigskip\endinsert\global\advance\figno by1}
\def\endinsert{}
\begin{document}

%\baselineskip 24pt

%%%%%%%%%%%%%%%%%%%%%%%%%%%%%%%%%%%%%%%%%%%%%%%%%%%%%%%%%%%%%%%%%
%%%
%%%                      TITLE PAGE
%%%
%%%%%%%%%%%%%%%%%%%%%%%%%%%%%%%%%%%%%%%%%%%%%%%%%%%%%%%%%%%%%%%%%

\title{\large{\bf Cosmology of Nonlinear Oscillations}}

\author{Stephen D.H.~Hsu\thanks{hsu@duende.uoregon.edu} \\
\\
Department of Physics, \\
University of Oregon, Eugene OR 97403-5203 \\ \\   }

%\date{January 21, 2000}

%\hfill{\parbox[b]{1in}{\hbox{\tt OITS
%711} }}

\maketitle

\begin{picture}(0,0)(0,0)
\put(350,365){\tt OITS-731}
%\put(350,350){\tt draft \today}
\end{picture}
\vspace{-24pt}

\begin{abstract}
The nonlinear oscillations of a scalar field are shown to have
cosmological equations of state with $w = p / \rho$ ranging from
$-1 < w < 1$. We investigate the possibility that the dark energy
is due to such oscillations.
\end{abstract}

%%%%%%%%%%%%%%%%%%%%%%%%%%%%%%%%%%%%%%%%%%%%%%%%%%%%%%%%%%%%%%%%%
%%%
%%%                     INTRODUCTION
%%%
%%%%%%%%%%%%%%%%%%%%%%%%%%%%%%%%%%%%%%%%%%%%%%%%%%%%%%%%%%%%%%%%%

\newpage

Astrophysical data support the existence of dark energy
\cite{WMAP,PR}. Since many proposed solutions of the cosmological
constant problem lead to exactly zero vacuum energy for empty
space, it is natural to consider so-called quintessence models in
which the dark energy is comprised of some scalar field which is
slowly evolving towards its minimum \cite{PR}. The main objections
to these models are their typically unnatural potentials, and that
they require the suppression of higher dimension operators likely
to be induced by quantum gravity \cite{sc}.

In this letter we investigate a qualitatively different idea: that
the dark energy is due to (possibly rapid) nonlinear oscillations
rather than slow evolution on cosmological timescales. We consider
oscillations in scalar models
\begin{equation}
{\cal L} = {1 \over 2} (\partial_{\mu} z)^2  - V(z)~,
\end{equation}
where the potential $V(z) = a \vert z \vert^l$ near its minimum.
Potentials with $l < 2$ are particularly interesting, as we will
see below that they yield an equation of state $w = p / \rho < 0$.
This form of $V(z)$ may appear odd, but a change of variables to,
e.g., $\phi = \vert z \vert^{l/2}$ yields
\begin{equation}
\label{K} {\cal L} = {1 \over 2} K(\phi) (\partial_{\mu} \phi)^2 -
a \phi^2~,
\end{equation}
with $K(\phi) = \left( 2 \over l \right)^2 \phi^{4- 2l \over l}$.
A kinetic term of this type can be obtained from the K\"ahler
potential in supersymmetric models. We focus here on the classical
behavior of $z$, but its quantization when $l < 2$ warrants
further investigation. It has the unusual property that there are
no perturbative degrees of freedom - that is, small oscillations
about $z=0$ have infinite frequency, since $V''(z=0)$ does not
exist. Only large (non-infinitesimal) $z$ oscillations can have
finite energy density. This may lead to a number of interesting
features: one might expect that $z$ decay and production rates, as
well as radiative corrections, only arise from nonperturbative
effects and are exponentially small.

The redshift of a field undergoing nonlinear oscillations can be
calculated through its average equation of state and depends on
the ratio $w = p/ \rho$. From the scalar field equations in a
Robertson-Walker universe, one obtains
\begin{equation}
\label{redshift} \rho (t) = \rho_0 \left( R_0 \over R(t)
\right)^{3(1+w)}~~~,
\end{equation}
where $R$ is the Robertson-Walker scale factor. The scalar field
energy redshifts like radiation when $w = 1/3$, like matter when
$w=0$, etc.

It remains to calculate the equation of state for nonlinear
oscillations. We note that $p = T - V$ and $\rho = T + V$, where
$T$ is the kinetic energy density and $V$ is the potential energy
density. We calculate the relation between T and V averaged over
an oscillation period, which is much smaller than the cosmological
timescale except during the very first oscillations, which begin
when the age of the universe is of order ${V'' (z)\,}^{-1/2}$.

We define
\begin{equation}
\langle T \rangle = \frac{1}{2} \int dt~\dot{z}^2
\end{equation}
and
\begin{equation}
\langle V \rangle = \int dt~ {z^{l}}~~~,
\end{equation}
where each integral is taken over the same period with boundary
conditions $\dot{z} (0) = \dot{z} (\tau) = 0$ (or equivalently $z
= 0$ at the endpoints) and we have adopted units in which the
overall scale of the potential is unity. Since
\begin{equation}
\frac{d}{dt} \left( z \dot{z} \right) = \dot{z}^2 + z \ddot{z}~~~,
\end{equation}
and the equation of motion is $\ddot{z} = - l \, z^{l-1}$, we can
rewrite the average potential energy as
\begin{equation}
\langle V \rangle = {1 \over l} \int dt ~\dot{z}^2 = {2 \over l}
\langle T \rangle~~~.
\end{equation}
This yields
\begin{equation}
\label{w} w = { (l-2) \over (l+2)}~~~,
\end{equation}
with $-1 < w < 1$. In potentials with $l < 2$, the average
potential energy dominates over the kinetic part, and the pressure
is negative. In the limit $l \rightarrow 0$ these oscillations
behave like a cosmological constant. In higher order potentials
$(l > 2)$, the situation is reversed, leading asymptotically to
$w=1$ as $l \rightarrow \infty$ , or $\rho \sim 1 / R^6$. These
oscillations redshift away rapidly, although it was noted in
\cite{Liddle} that the large l behavior of $w = 1$ can never be
achieved, due to an instability to a non-oscillatory scaling
solution.

Given a periodic solution to the $z$ equations of motion, one can
obtain a rescaled solution via
\begin{equation}
z(t) \rightarrow ~a^{2 \over l-2} ~z (at)~~.
\end{equation}
For $l < 2$, the frequency of oscillation goes to infinity as the
amplitude goes to zero. Note, however, that the average energy
density goes to zero in this limit.

An advantage of nonlinear oscillation models of quintessence is
that the potential $V(z)$ need not be characterized by the size of
the current dark energy density $\sim (10^{-3} \, {\rm eV})^4$,
nor be fine-tuned to be flat (i.e., have curvature of order the
inverse horizon size squared $\sim (10^{-33} \, {\rm eV})^2$).
Rather, the potential can be characterized by a larger energy
scale more familiar to particle physics, with no small
dimensionless parameters. The smallness of the energy density
today relative to this scale could be explained by a small $z$
oscillation amplitude.

One scenario that would result in a small oscillation amplitude is
if the original energy density in the $z$ field were diluted away
by inflation, and the subsequent reheat temperature insufficient
to repopulate it. This is quite plausible if the couplings between
the inflaton and ordinary matter fields to the $z$ field are
small, for example suppressed by the Planck scale if $z$ is a
hidden sector field\footnote{Some reheating of the $z$ field is
inevitable, even if its couplings to the inflaton are very small.
However, the thermal $z$ bosons produced do not necessarily
contribute to the coherent oscillations studied here - their
energy redshifts away more rapidly.}. The current $z$ energy
density is dependent on the number of e-foldings N during
inflation
\begin{equation}
\rho_{\rm today} \sim ~\rho_i~ \left[ 2 \cdot 10^{-5} \, e^{-N}
\left( T_{\rm md} \over T_{\rm rh} \right)\right]^{\nu}~~,
\end{equation}
where $\nu = 3(1+w)$, $T_{\rm md} \sim 5 ~ {\rm eV}$ is the
temperature at which the universe becomes matter dominated and
$T_{\rm rh}$ is the reheat temperature after inflation. For
example, using $\nu = 1/2$ (or $w = -5/6$, consistent with the
WMAP limit of $w < -0.78$ \cite{WMAP}), $T_{\rm rh} = 5 \cdot
10^{10} ~{\rm GeV}$ and $\rho_i$ the Planck energy density, we
find that $N \sim 510$ in order that $\rho_{\rm today} \sim
(10^{-3} \, {\rm eV})^4$. For smaller $\rho_i \sim (10^{11} \,
{\rm GeV})^4$, appropriate for intermediate scale inflation, we
find $N \sim 370$. Note, while the scenario described here
explains the small dark energy density today, it does not address
the question of coincidence: why is $\rho_z$ of order $\rho_{\rm
critical}$ today?

If the energy scale characterizing $V(z)$ is small (within several
orders of magnitude of an electron volt) little dilution is
necessary, and it could be provided by the expansion of the
universe after inflation. If $z$ originates from the scalar
component of a chiral superfield $\Phi$, the energy scale of
$V(z)$ (i.e., the parameter $a$ in (\ref{K})) is protected by
supersymmetry from radiative corrections. A Yukawa coupling
between $z$ and its superpartner can be excluded by imposing $\Phi
\rightarrow - \Phi$ symmetry in the superpotential, thereby
stabilizing $z$.

To conclude, we find that realistic quintessence models based on
nonlinear oscillations can be constructed without any fine-tuning
of fundamental parameters. These models will be disfavored if
future data show that $w = -1$.

\bigskip
\noindent {\bf Note added:} After this work was completed, we
learned that the result (\ref{w}) for the equation of state was
previously derived by M. Turner in \cite{Turner}. Also, models
based on a hyperbolic cosine potential raised to a fractional
power have been considered in \cite{cosh}, which at late times
exhibit oscillations of the type considered here.

%\bigskip
\newpage

%%%%%%%%%%%%%%%%%%%%%%%%%%%%%%%%%%%%%%%%%%%%%%%%%%%%%%%%%%%%%%%%%
%%%
%%%                   ACKNOWLEDGEMENTS
%%%
%%%%%%%%%%%%%%%%%%%%%%%%%%%%%%%%%%%%%%%%%%%%%%%%%%%%%%%%%%%%%%%%%
\section*{Acknowledgements}
\noindent We would like to thank D.K. Hong and A. Zee for useful
comments. This work was supported in part under DOE contract
DE-FG06-85ER40224.

%%%%%%%%%%%%%%%%%%%%%%%%%%%%%%%%%%%%%%%%%%%%%%%%%%%%%%%%%%%%%%%%%
%%%
%%%                     BIBLIOGRAPHY
%%%
%%%%%%%%%%%%%%%%%%%%%%%%%%%%%%%%%%%%%%%%%%%%%%%%%%%%%%%%%%%%%%%%%

\bigskip

%\newpage

\end{document}